\documentclass[11pt]{article}

\usepackage[final]{acl}

\usepackage{times}
\usepackage{latexsym}

\usepackage[T1]{fontenc}
\usepackage[utf8]{inputenc}

\usepackage{microtype}

\usepackage{inconsolata}

\usepackage{graphicx}
\usepackage{multirow}
\usepackage{makecell}
\usepackage{placeins}
\usepackage{amssymb}
\usepackage{booktabs}
\usepackage[most]{tcolorbox}
\tcbuselibrary{listings, breakable}
\usepackage{xcolor}
\usepackage{colortbl}
\usepackage{graphicx}
\usepackage{enumitem}
\usepackage{caption}
\usepackage{amsmath, amssymb}

\usepackage{booktabs}
\usepackage{tabularx}
\usepackage{adjustbox}
\usepackage{placeins}
\usepackage{algorithm}
\usepackage{algpseudocode}
\title{
ToxReason: 
A Benchmark for Mechanistic Chemical Toxicity Reasoning via Adverse Outcome Pathway
}
\author{
 \textbf{Jueon Park\textsuperscript{1}}, \
 \textbf{Wonjune Jang\textsuperscript{2}}, \
 \textbf{Chanhwi Kim\textsuperscript{3}}, \
 \textbf{Yein Park\textsuperscript{1,4}}, \
 \textbf{Jaewoo Kang\textsuperscript{1,4,$\dagger$}}
\\
 \textsuperscript{1}Korea University \
 \textsuperscript{2}Myongji University \\ 
 \textsuperscript{3}University of Texas Health Science Center at Houston \ 
 \textsuperscript{4}AIGEN Sciences
\\
\{jueon\_park, kangj\}@korea.ac.kr
}

\usepackage{natbib}
\usepackage{url}
\begin{document}
\maketitle
\begin{abstract}
Recent advances in large language models (LLMs) have enabled molecular reasoning for property prediction. 
However, toxicity arises from complex biological mechanisms beyond chemical structure, necessitating mechanistic reasoning for reliable prediction.
Despite its importance, current benchmarks fail to systematically evaluate this capability. LLMs can generate fluent but biologically unfaithful explanations, making it difficult to assess whether predicted toxicities are grounded in valid mechanisms.
To bridge this gap, we introduce ToxReason, a benchmark grounded in the Adverse Outcome Pathway (AOP) that evaluates organ-level toxicity reasoning across multiple organs. 
ToxReason integrates experimental drug–target interaction evidence with toxicity labels, requiring models to infer both toxic outcomes and their underlying mechanisms from Molecular Initiating Event (MIE) to Adverse Outcome (AO).
Using ToxReason, we evaluate toxicity prediction performance and reasoning quality across diverse LLMs.
We find that strong predictive performance does not necessarily imply reliable reasoning.
Furthermore, we show that reasoning-aware training improves mechanistic reasoning and, consequently, toxicity prediction performance. 
Together, these results underscore the necessity of integrating reasoning into both evaluation and training for trustworthy toxicity modeling.

\end{abstract}
\section{Introduction}

\begin{figure}[t]
  \centering
  \includegraphics[width=\linewidth]{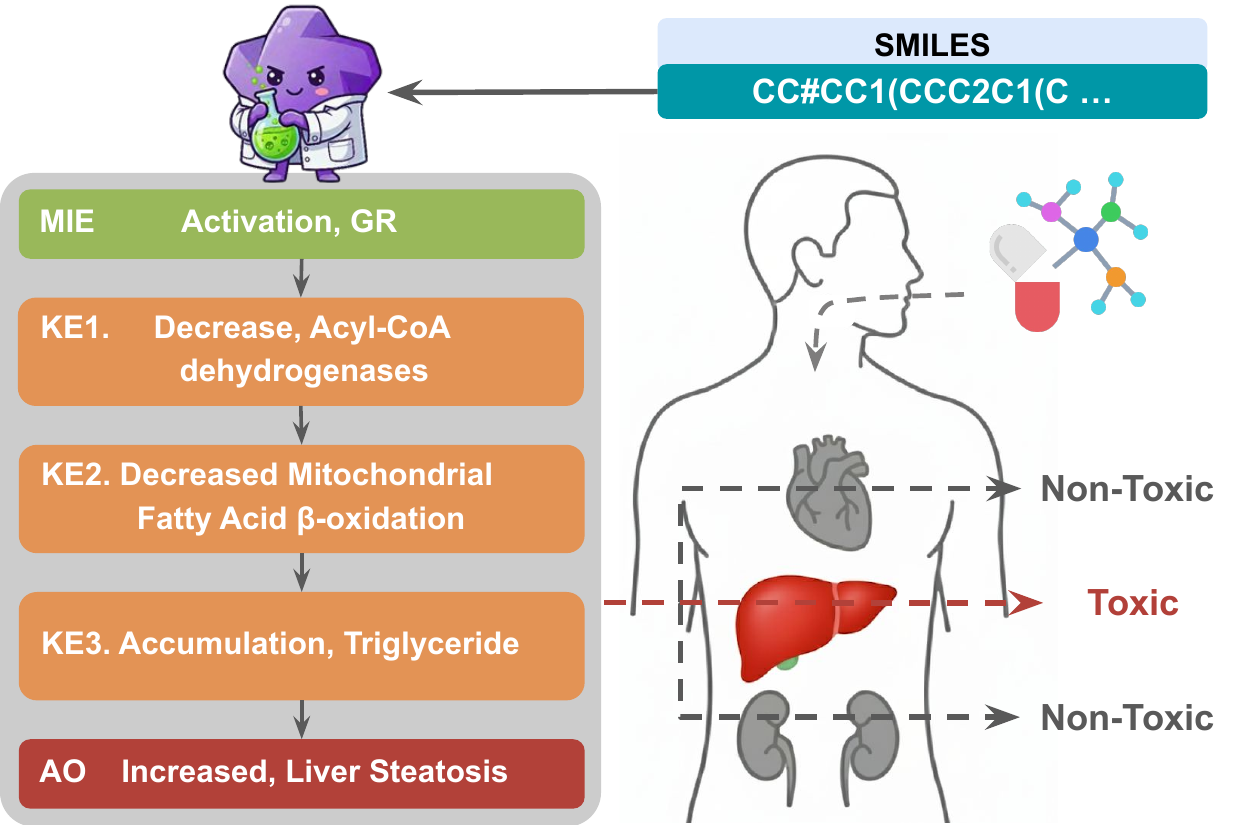}
  \vspace{-20pt}
  \caption{An example of AOP-based mechanistic toxicity reasoning from a MIE to an organ-level AO.}
  \vspace{-16pt}
  \label{fig:toxreason_preview_figure}
\end{figure}

Large language models (LLMs) have recently been applied to molecular reasoning tasks, allowing property prediction directly from chemical representations such as SMILES~\citep{weininger1988smiles}. Prior works~\citep{jang2025structural, kim2025mol, zhuang2025reasoning} have shown that LLMs can capture molecular structure and semantics, with explicit reasoning further improving predictive performance. Correspondingly, several benchmarks~\citep{liu2025fgbench, li2025beyond} have been proposed to evaluate molecular reasoning abilities, primarily focusing on structure–property relationships. However, toxicity represents a fundamentally different challenge, as toxic effects often arise from complex biological mechanisms.
These involve molecular targets, downstream cellular events, and organ-level responses, rather than chemical structure alone~\citep{xu2020systematic, uesawa2024efficiency}.

In toxicology, such mechanistic processes are systematically described by the Adverse Outcome Pathway (AOP) framework. It represents toxicity as a causal sequence from a molecular initiating event (MIE) to downstream key events (KE) and ultimately an adverse outcome (AO) at the organ level~\citep{leist2017adverse, vinken2017adverse}. As illustrated in Figure~\ref{fig:toxreason_preview_figure}, a query molecule may first activate the glucocorticoid receptor (GR), which leads to decreased activity of acyl-CoA dehydrogenases and decreased mitochondrial fatty acid $\beta$-oxidation. 
These disruptions result in triglyceride accumulation and ultimately leading to liver steatosis, an AO and a key manifestation of liver toxicity. Notably, this structure closely aligns with multi-step reasoning commonly studied in natural language processing, where complex conclusions are derived through intermediate reasoning steps~\citep{wei2022chain, wang2022self, hwang2025assessing}.

Despite the importance of mechanistic reasoning in toxicity, existing datasets(e.g., Tox21~\cite{huang2016tox21challenge} and ClinTox~\cite{gayvert2016data}) have not been designed to evaluate whether models can reason about toxicity through biologically grounded mechanisms. While UniTox~\citep{silberg2024unitox} derives toxicity labels and corresponding explanations by summarizing clinical evidence from openFDA documents, its reasoning is primarily grounded in observed adverse effects rather than biologically mechanistic pathways.
We propose ToxReason, a novel benchmark designed to evaluate mechanistic toxicity reasoning grounded in biologically causal processes. It requires LLMs to infer toxicity through structured reasoning over molecular interactions and downstream biological events. Evaluating this capability is particularly important in scenarios where clinical observations are unavailable, such as early-stage drug discovery and chemical safety assessment~\citep{zheng2025large}.

ToxReason is constructed by integrating structured knowledge of causal toxicity pathways with experimental drug–target interaction data and curated chemical–toxicity associations. This yields a scientifically grounded benchmark of high-fidelity reasoning instances for 193 chemicals. Using this benchmark, we assess whether LLMs can move beyond predicting toxic outcomes to reasoning about underlying toxicity mechanisms. We employ an LLM-based evaluator to assess the reasoning quality based on four complementary metrics. Our analysis reveals that predictive performance and reasoning do not always align across models, suggesting differences in how LLMs internalize toxicity-related knowledge. This misalignment raises concerns about the reliability of toxicity predictions when models achieve high accuracy without corresponding mechanistic reasoning.

Beyond evaluation, we construct training data to explore three distinct learning paradigms. Specifically, our approach using reinforcement learning, explicitly optimizes both toxicity prediction and reasoning. Through this approach, our compact 4B-parameter model achieves performance comparable to or exceeding state-of-the-art models in toxicity prediction, while also demonstrating substantially improved reasoning ability.
These results underscore the necessity of reasoning-aware optimization for aligning model predictions with biologically grounded explanations.

Our contributions are summarized as follows:
\begin{itemize}
    \item We introduce ToxReason, a mechanistic toxicity benchmark that combines drug toxicity labels with AOP-based causal reasoning, enabling evaluation beyond outcome prediction.
    \item We systematically evaluate multiple open- and closed-source LLMs in terms of how they reason over toxicological mechanisms, rather than relying solely on surface-level toxicity prediction.
    \item We show that explicitly learning mechanistic toxicity reasoning leads to a decisive improvement in toxicity prediction, allowing a compact model to outperform other larger state-of-the-art models.
\end{itemize}
\section{Related Work}

\paragraph{Adverse Outcome Pathway(AOP)}
The AOP is a conceptual framework that portrays the sequential chain of causal events across different levels of biological organization~\citep{zilliacus2024building}. It begins with a Molecular Initiating Event (MIE), which is the initial interaction between a chemical and a specific biological molecule like a protein or receptor. This trigger sets off a series of Key Events (KE) which are measurable biological changes that occur at the cellular, tissue, or organ level. These events act as "dominoes" that eventually lead to the Adverse Outcome (AO) such as toxicities. 

AOPs are typically developed through systematic integration of experimental evidence, literature curation, and expert knowledge to establish causal relationships~\citep{ankley2010adverse}. This mechanistic perspective has become central in modern toxicology, particularly in the new approach methodologies (NAMs) aimed at improving chemical risk assessment while reducing reliance on animal experimentation~\citep{saarimaki2023curated}.

\paragraph{LLMs for toxicity prediction}
Recent studies have explored large language models (LLMs) as tools for molecular toxicity prediction by leveraging their ability to reason over chemical structures and textual knowledge~\citep{zhang2025computational}. Previous work~\citep{yang2025large, chen2025application} demonstrated the feasibility of applying LLMs to specific toxicity endpoints, such as cardiotoxicity and drug-induced osteotoxicity. They primarily focus on predicting toxic outcomes from molecular representations such as SMILES. While these approaches showed promising predictive performance, they largely treated toxicity as an outcome prediction task and provided limited insight into the underlying biological mechanisms. 

More recently, CoTox~\citep{park2025cotox} introduced a Chain-of-Thought (CoT) framework that incorporates biological pathway and gene ontology information with structural context to generate toxicity reasoning, representing a step toward more interpretable predictions. However, this approach focuses on improving prediction capabilities by constructing explanatory narratives from given chemical and biological information, rather than explicitly assessing whether the reasoning process itself aligns with mechanistically grounded causal pathways. In contrast, ToxReason reframes mechanistic toxicity reasoning as an evaluation problem by providing a benchmark grounded in causal toxicity pathways.
This benchmark enables systematic evaluation of both LLMs’ toxicity prediction performance and the alignment of their reasoning with AOP.
\begin{figure*}[t]
  \centering
  \includegraphics[width=0.98\textwidth]{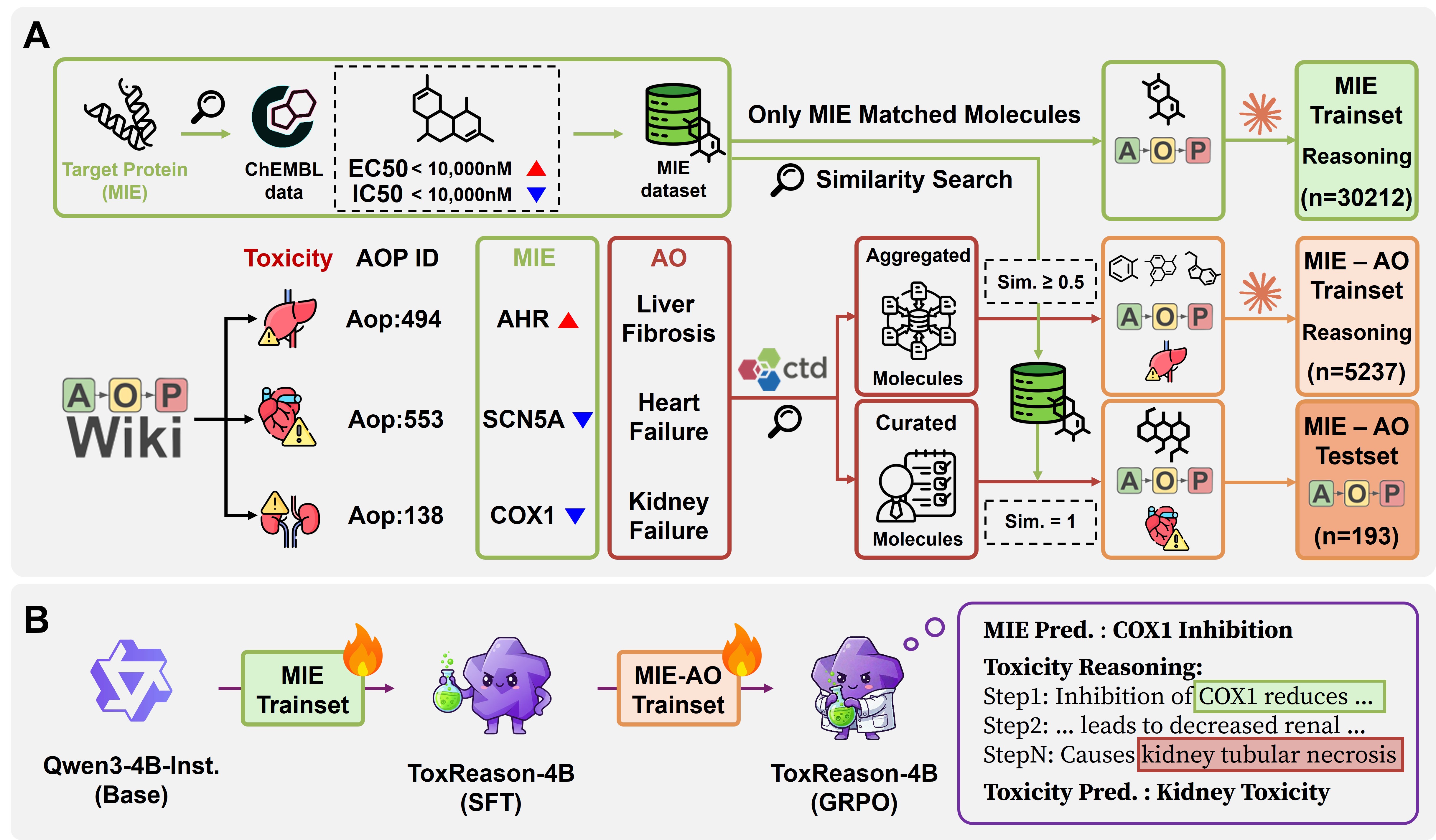}
  \caption{\textbf{Overview of ToxReason}, (A) Organ-specific AOPs are selected from AOP-Wiki, disease-linked chemicals are retrieved from CTD, and ChEMBL-derived MIE data are used to construct training and test sets under similarity constraints.(\textcolor{red}{$\blacktriangle$} Activation, \textcolor{blue}{$\blacktriangledown$} Inhibition) (B) Learning framework built on ToxReason, combining supervised fine-tuning and reinforcement learning for mechanistic toxicity reasoning.}
  \label{fig:main_figure}
\end{figure*}
\section{ToxReason Benchmark}
\subsection{Dataset Construction}
Figure~\ref{fig:main_figure} provides an overview of the ToxReason dataset construction pipeline.

\subsubsection{AOP Selection}
We first curated a set of AOPs focused on liver, heart, and kidney toxicities to build our mechanistic reasoning benchmark. These specific organ systems were selected because they are primary targets of drug-induced toxicity~\cite{rana2020most} and are well-documented in established AOP databases. AOPs were selected from AOP-Wiki\footnote{\url{https://aopwiki.org}}database (Release 2.7) by focusing on pathways whose AOs correspond to clinically meaningful organ-level toxicities and whose MIEs involve explicit activation or inhibition of specific protein targets. Based on this selection strategy, we identified 23 unique AOPs and curated 25 distinct molecular targets involved in MIEs. 
The full list of selected AOPs, along with their associated MIE targets and AOs,
is provided in Table~\ref{tab:aop_selection}.
These targets represent key biological entry points through which chemical perturbations can initiate toxicity pathways, providing a structured foundation for mechanistic toxicity reasoning in ToxReason.

\subsubsection{Chemical–AOP Association Derivation}
\label{sec:Chemical–AOP}
To derive mechanistic associations between chemicals and AOPs, we integrated disease–chemical relationships with experimental and similarity-based evidence of MIEs. For each selected AOP, its AO was treated as a disease concept(e.g., liver fibrosis and heart failure) and used to retrieve associated chemicals from the Comparative Toxicogenomics Database (CTD, Nov. 2025 release)~\citep{davis2025ctd}. These chemicals were considered candidate compounds linked to the corresponding AOP.

In parallel, we collected experimental evidence for MIE target proteins from ChEMBL (v36)~\citep{zdrazil2024chembl}. For each target, activity data were extracted based on assay type and direction of action, where EC50 values were used for activation and IC50 values for inhibition. A chemical was considered to activate or inhibit a target if the corresponding activity value was below 10,000 nM, following established criteria used in \citealt{gadaleta2024quantitative} on quantitative structure–activity relationships for MIEs of organ-specific toxicity. Based on these rules, we curated a unified MIE dataset by merging molecules annotated with their corresponding targets and directions of action across all selected MIEs.

To infer MIEs for each candidate chemical, we performed similarity-based evidence aggregation. Specifically, chemicals retrieved from CTD were treated as query molecules, and structurally similar compounds with known target activity were identified from MIE dataset. The direction of interaction (activation or inhibition) for each MIE target was determined by majority voting over the activity annotations of these similar compounds. This procedure enabled robust inference of MIEs even when direct experimental measurements for the query chemical were unavailable.

Finally, we associated each chemical with an AOP when its inferred MIE and observed AO aligned within the same pathway, allowing the resulting AOP to function as the mechanistic toxicity rationale for the chemical.

\subsubsection{Construct Train and Testset}
To support both learning and evaluation of mechanistic toxicity reasoning, we constructed separate train and test sets under different evidence conditions, each serving a distinct role in the overall benchmark design. Detailed data statistics are provided in Appendix~\ref{sec:appendix_details}.
\paragraph{Training Data}
The training data were constructed as two complementary sets, namely the MIE-matched trainset and the MIE-AO-matched trainset. For the MIE-matched trainset, we collected compounds that satisfy the MIE condition of a given AOP based on experimental evidence of target-specific activation or inhibition. These compounds were matched to an AOP through their associated MIEs without requiring confirmation of the corresponding AO. Each compound in this set corresponds to a case where the toxic mechanism is known to have been initiated. While training, the MIE-matched trainset helps the model robustly learn AOP-specific initiation patterns by increasing coverage of MIE-matched compounds.

For the MIE-AO-matched trainset, we retained only compounds that satisfy both the MIE condition and the corresponding AO evidence for the same AOP. This trainset was constructed using the Chemical–AOP association derivation procedure described in Section \ref{sec:Chemical–AOP}. Here, aggregated CTD associations, which include both curated and inference-based chemical–disease relationships, were used to obtain disease-level toxicity labels, and similarity-based inference was applied with a relaxed Dice similarity~\citep{dice1945measures} threshold ($\geq 0.5$) to infer MIEs from structurally related compounds. This setting encourages the model to reason across molecular interactions and downstream toxic outcomes.

For both trainsets, mechanistic toxicity explanations were generated in an instruction-style format using Claude Haiku 4.5~\citep{anthropic2025claudehaiku45} model. Given the inferred AOP structure, the model was prompted to narratively describe why a specific chemical could induce a particular toxicity through MIEs and subsequent biological processes.

\paragraph{Test Data}
The test set was constructed to evaluate mechanistic toxicity reasoning under strictly controlled evidence conditions. To avoid confounding effects from inferred or indirect associations, we relied on curated, human-specific CTD associations to define AOs. In addition, similarity-based inference was restricted to cases where the query compound was structurally identical to reference compounds, thereby eliminating uncertainty arising from structural approximation. As a result, each test instance contains the molecular structure, the corresponding MIE annotations, and the associated AOP context used for mechanistic reasoning. All molecules in the test set were strictly excluded from both training sets to prevent any overlap and ensure an unbiased evaluation.

\subsection{Task Definition}
ToxReason defines a mechanistic toxicity reasoning task that evaluates whether a model can infer toxic outcomes through biologically grounded reasoning rather than surface-level prediction. Given a query molecule, the task requires models to reason over molecular interactions and downstream biological processes to explain how toxicity manifests in a step-by-step manner.

Specifically, the task begins with a query chemical accompanied by contextual evidence retrieved from structurally similar compounds, including their similarity scores and experimentally supported MIE signals. This formulation is inspired by MolRAG~\citep{xian2025molrag}, which leverages outcomes from structurally related molecules to support property inference for a query compound. Following this paradigm, we retrieve evidence from a ChEMBL-derived MIE dataset by collecting experimentally supported MIE outcomes from compounds that are structurally similar to the query molecule. We provide four activation and four inhibition examples as contextual evidence, following the four-shot setting adopted in MolRAG.

Based on the predicted MIEs, models then perform step-wise mechanistic reasoning, tracing how each molecular interaction may propagate through downstream biological events and ultimately lead to organ-level toxicity. The final output consists of a predicted toxic outcome along with a mechanistic explanation that connects MIEs to AOs through biologically meaningful reasoning steps.
\begin{table*}[t]
\centering
\small
\setlength{\tabcolsep}{5pt}
\renewcommand{\arraystretch}{1.15}

\resizebox{\textwidth}{!}{%
\begin{tabular}{lcccccccccc}
\noalign{\hrule height 1.5pt}
\multicolumn{1}{c}{\multirow{2}{*}[-0.35ex]{Models}}
& \multicolumn{4}{c}{\rule{0pt}{2.9ex}Toxicity Prediction (F1-Score, \%)}
& \multicolumn{4}{c}{\rule{0pt}{2.9ex}Toxicity Reasoning (LLM-as-a-Judge)}
& \multicolumn{1}{c}{\rule{0pt}{2.9ex}{NW}} \\
\cmidrule(lr){2-5} \cmidrule(lr){6-9} \cmidrule(lr){10-10}
& \rule{0pt}{1.6ex}Kidney & Cardio & Liver & Avg.
& \rule{0pt}{1.6ex}Halluc. & Causal. & Biol. & Overall
& \rule{0pt}{1.6ex}{Score} \\
\hline

\textit{\textbf{Closed Model}} & & & & & & & & & \\
GPT-4o      & 54.8 & 74.6 & 60.2 & 63.2 & 4.933 & 5.244 & 5.311 & 4.959 & 0.494 \\
GPT-5       & 56.4 & 72.7 & \underline{65.0} & 64.7 & \underline{5.627} & 5.788 & \underline{6.176} & 5.420 & 0.475 \\
GPT-5.1     & 50.3 & 71.2 & 58.9 & 60.1 & 5.611 & \underline{5.881} & \textbf{6.378} & \underline{5.523} & 0.513 \\
o3          & 60.0 & 72.5 & 58.8 & 63.8 & 5.342 & 5.736 & 5.922 & 5.326 & 0.488 \\
o4-mini     & 58.5 & 71.3 & 55.6 & 61.8 & 4.808 & 5.549 & 5.492 & 4.948 & 0.487 \\

\hline
\textit{\textbf{Open Model}} & & & & & & & & & \\
Qwen3-4B-Inst.        & 56.9 & 71.1 & 57.3 & 61.8 & 4.337 & 4.969 & 5.104 & 4.523 & 0.484 \\
Llama3.1-8B-Inst.    & 31.2 & 60.2 & 55.7 & 49.0 & 3.269 & 3.689 & 3.736 & 3.275 & 0.396 \\
Qwen2.5-14B-Inst.    & 48.8 & 65.0 & 58.3 & 57.4 & 4.492 & 4.767 & 5.000 & 4.528 & 0.484 \\
Gemma3-27B-Inst.     & 54.2 & 69.1 & 56.6 & 60.0 & 4.876 & 5.326 & 5.451 & 4.959 & 0.498 \\
Llama3.1-70B-Inst.   & 57.1 & 76.1 & 57.4 & 63.5 & 4.554 & 4.865 & 5.016 & 4.653 & 0.473 \\
DeepSeek-R1-Distill-70B  
& 59.1 & \textbf{78.5} & 59.6 & 65.7 & 4.508 & 4.679 & 4.876 & 4.487 & 0.458 \\

\hline
\textit{\textbf{In-Context Learning}} & & & & & & & & & \\
Qwen3-4B-ICL-1shot & \underline{68.4} & \underline{77.7} & 60.3 & \underline{68.8} & 5.259 & 5.694 & 5.922 & 5.373 & \underline{0.519} \\
Qwen3-4B-ICL-2shot & 55.2 & 63.5 & 58.7 & 59.1 & 4.233 & 5.073 & 5.223 & 4.373 & 0.404 \\
Qwen3-4B-ICL-4shot & 33.7 & 70.9 & 59.3 & 54.6 & 4.145 & 4.762 & 4.990 & 4.212 & 0.418 \\

\hline
\textit{\textbf{Supervised Finetuning}} & & & & & & & & & \\
ToxReason-4B-SFT & 57.9 & 74.3 & 57.4 & 63.2 & 4.399 & 4.927 & 5.166 & 4.554 & 0.481 \\

\hline
\textit{\textbf{Reinforcement Learning}} & & & & & & & & & \\
ToxReason-4B-GRPO 
& \textbf{73.4} & 72.7 & \textbf{68.2} & \textbf{71.4} 
& \textbf{5.725} & \textbf{5.896} & 5.642 & \textbf{5.642} & \textbf{0.720} \\

\noalign{\hrule height 1.5pt}
\end{tabular}
}
\vspace{-5pt}
\caption{Performance comparison of toxicity prediction (0--100), mechanistic toxicity reasoning (0--10), and NW alignment (0--1) across models. \textbf{Bold} indicates the best score and \underline{underline} indicates the second-best score.
}
\label{tab:main_results}
\vspace{-7pt}
\end{table*}

\subsection{Evaluation}
ToxReason evaluates LLMs from two complementary perspectives, focusing on toxicity prediction and reasoning. This dual evaluation is designed to distinguish models that predict toxic outcomes from those that can also explain toxicity through biologically grounded mechanisms.

\paragraph{Toxicity Prediction}
For toxicity prediction, models are evaluated on a multi-label classification task covering three organ-level toxicities: liver, heart, and kidney toxicity. Given a query chemical, models are instructed to output the names of predicted toxicities in natural language. These responses are then post-processed into binary labels (0/1) for each organ, indicating the absence or presence of toxicity. Performance is measured using the F1-score for each toxicity type, along with the macro-averaged F1-score across all three toxicities, providing a balanced assessment of predictive performance.

\paragraph{Toxicity Reasoning}
To assess mechanistic reasoning quality, we compare model-generated explanations against the reference AOPs associated with each query chemical. Given the open-ended nature of natural language explanations, we adopt an LLM-as-a-Judge~\citep{zheng2023judging} framework to evaluate reasoning quality. Specifically, Claude Sonnet 4.5~\citep{anthropic2025claudesonnet45} is used as an independent evaluator to score each response by jointly considering the model's response and the corresponding ground-truth AOP.

Reasoning quality is evaluated along the following dimensions, each scored on a 0–10 scale:
\begin{itemize}
    \item \textbf{Hallucination Avoidance} measures the extent to which the explanation avoids introducing unsupported or fabricated information.
    \item \textbf{Causal Coherence} evaluates whether the mechanistic reasoning follows a logically consistent causal chain, ensuring that MIEs lead to KEs and AOs in the correct order without contradictions or unjustified transitions.
    \item \textbf{Biological Fidelity} assesses the biological validity of the explanation, including correct use of toxicological terminology, accurate relationships between MIEs, KEs, and AOs, and consistency with known liver, heart, and kidney toxicology.
    \item \textbf{Overall} provides a holistic assessment of the explanation, summarizing the model’s overall ability to produce coherent and biologically grounded mechanistic reasoning.
\end{itemize}

Scores for each criterion are reported separately, enabling fine-grained analysis of reasoning behavior across models, in addition to an overall reasoning quality score. Detailed prompts can be found in Table \ref{tab:llm_as_judge_system_prompt} and \ref{tab:llm_as_judge_user_prompt}.

\section{Experiment Setup}

\subsection{Prompt Design}
Models are prompted to infer MIEs for a query chemical based on experimental evidence retrieved from structurally similar compounds. For each inferred MIE, models generate a distinct step-wise mechanistic explanation that links the specific event to its corresponding organ-level toxicity. The model produces a structured JSON output including MIE predictions, individualized mechanistic reasoning for each event, and a summary of the final predicted toxicities across the three target organs. Prompts used in our experiments are provided in the Table \ref{tab:llm_inference_system_prompt} and \ref{tab:llm_inference_user_prompt}.

\subsection{Models}
We evaluate how different LLMs perform on the ToxReason benchmark under a unified zero-shot setting. The evaluated models include closed-systems such as GPT-4o~\cite{hurst2024gpt}, GPT-5~\cite{openai_gpt5_system_card}, GPT-5.1~\cite{openai_gpt5_1_system_card}, o3, and o4-mini~\cite{openai_o3_o4mini_system_card}, as well as open models including Qwen2.5-14B-Instruct~\cite{yang2024qwen2}, Llama3.1-8B-Instruct, Llama3.1-70B-Instruct~\cite{grattafiori2024llama}, Qwen3-4B-Instruct~\cite{yang2025qwen3}, Gemma3-27B-Instruct~\cite{team2025gemma}, and Deepseek-R1(Llama-70B Distilled)~\cite{guo2025deepseek} covering a broad range of model scales. All models are prompted using the same task formulation and evaluated with identical output processing and scoring procedures. In addition, all experiments employ a greedy decoding strategy with a temperature of 0 and top\_p set to 1.0 to ensure consistent and deterministic inference across models.

\subsection{Model Improvement Strategies}
To investigate if ToxReason-derived train data can effectively improve model performance, we evaluate three learning paradigms using Qwen3-4B-Instruct as base model. We first examine in-context learning~\citep{brown2020language} to assess the impact of few-shot demonstrations provided at inference time. Following this, we perform supervised fine-tuning \citep{zhang2023instruction} through LoRA-based adaptation to align the model with the ToxReason task structure. Finally, we implement a two-stage reinforcement learning framework using Group Relative Policy Optimization (GRPO)~\citep{shao2024deepseekmath} to explicitly optimize for AOP-grounded reasoning and causal consistency.  The full implementation details for each of these three learning strategies are provided in Appendix~\ref{sec:appendix_training_details}.

\section{Results and Analysis}

\subsection{Zero-shot Performance Comparison}
As shown in Table \ref{tab:main_results}, closed models tend to exhibit stronger mechanistic reasoning performance, while predictive performance is more comparable between closed and open models. Especially, DeepSeek-R1 and GPT-5 achieved the strongest predictive performance, GPT-5.1 demonstrated the best reasoning quality, as evidenced by its overall score of 5.523. However, GPT-5.1 exhibited the lowest predictive performance among all evaluated closed models at 60.1\%. This result illustrates a distinct gap between predicting toxic labels and the ability to provide mechanistic reasoning.

Among open models, DeepSeek-R1 delivered the strongest predictive performance, achieving the highest cardiotoxicity F1-score across all models. However, its mechanistic reasoning performance remained relatively limited, comparable to or lower than that of smaller open models, which may reflect its design emphasis on broad, task-agnostic reasoning rather than biology-grounded mechanistic understanding.

These findings confirm a significant misalignment between predictive results and reasoning quality. This discrepancy highlights the necessity of ToxReason for evaluating a model's reasoning process instead of relying on predictive outcomes.

\subsection{Effectiveness of Improvement Strategies}
Applying various learning strategies to the Qwen3-4B base model revealed distinct performance patterns across the benchmark. In-context learning (ICL) achieved its peak effectiveness in a 1-shot configuration, significantly improving the average predictive performance to 68.8\% and reasoning quality to 5.373. However, both performances declined as the shot count increased. This suggests that providing additional demonstrations may introduce contextual noise that hinders the model.

Additionally, supervised fine-tuning showed negligible differences in both predictive performance and reasoning quality compared to the base model. In contrast, reinforcement learning through the GRPO framework produced the most substantial gains by explicitly optimizing reasoning ability. The resulting ToxReason-4B-GRPO model attained an average predictive performance of 71.4\% and an overall reasoning score of 5.642, significantly outperforming the base model and surpassing even the most capable closed-models.

\begin{table}[t]
\large   
\centering

\setlength{\tabcolsep}{10pt}

\renewcommand{\arraystretch}{1.5}

\resizebox{1\linewidth}{!}{%

\begin{tabular}{lcc}

\noalign{\hrule height 1.5pt}

\multirow{2}{*}{\raisebox{-1ex}{\textbf{LLM-as-a-Judge}}}
& \multicolumn{2}{c}{\textbf{NW-alignment Score}} \\

\cmidrule(lr){2-3}

& \raisebox{0.3ex}{\textbf{Pearson $r$}}
& \raisebox{0.3ex}{\textbf{Spearman $\rho$}} \\

\hline
Halluc. Avoidance & 0.689 {\footnotesize ($p < 0.001$)} & 0.739 {\footnotesize ($p < 0.001$)} \\
Causal Coherence        & 0.615 {\footnotesize ($p < 0.01$)} & 0.739 {\footnotesize ($p < 0.001$)} \\
Biological Fidelity    & 0.457 {\footnotesize ($p = 0.055$)} & 0.692 {\footnotesize ($p < 0.001$)} \\
Overall                & 0.703 {\footnotesize ($p < 0.001$)} & 0.837 {\footnotesize ($p < 0.001$)} \\

% Hallucination Avoidance & 0.686 {\scriptsize ($p = 1.76 \times 10^{-3}$)} & 0.713 {\scriptsize ($p = 7.57 \times 10^{-4}$)} \\
% Causal Coherence        & 0.618 {\scriptsize ($p = 9.09 \times 10^{-3}$)} & 0.713 {\scriptsize ($p = 7.57 \times 10^{-4}$)} \\
% Biological Fidelity     & 0.447 {\scriptsize ($p = 9.77 \times 10^{-2}$)} & 0.631 {\scriptsize ($p = 7.00 \times 10^{-3}$)} \\
% Overall                 & 0.702 {\scriptsize ($p = 1.09 \times 10^{-3}$)} & 0.806 {\scriptsize ($p = 6.27 \times 10^{-6}$)} \\

% Hallucination Avoidance & 0.686 {\scriptsize ($p = 0.002$)} & 0.713 {\scriptsize ($p = 0.0008$)} \\
% Causal Coherence        & 0.618 {\scriptsize ($p = 0.009$)} & 0.713 {\scriptsize ($p = 0.0008$)} \\
% Biological Fidelity     & 0.447 {\scriptsize ($p = 0.098$)} & 0.631 {\scriptsize ($p = 0.007$)} \\
% Overall                 & 0.702 {\scriptsize ($p = 0.001$)} & 0.806 {\scriptsize ($p = 6.27 \times 10^{-6}$)} \\

\noalign{\hrule height 1.5pt}

\end{tabular}

}
\vspace{-5pt}
\caption{Correlation between NW-alignment scores and LLM-as-a-Judge toxicity reasoning metrics.
$p$-values denote statistical significance.}

\label{tab:nw_alignment}
\vspace{-10pt}
\end{table}

% \begin{figure}[t]
%   \centering
%   \includegraphics[width=\linewidth]{_image/corr_graph.png}
%   \caption{Correlation between reasoning quality assessed by an LLM-as-a-Judge and mechanistic path alignment quantified using the Needleman–Wunsch algorithm.}
%   \vspace{-10pt}
%   \label{fig:reasoning_analysis}
% \end{figure}

\subsection{Analysis of Toxicity Reasoning Metrics}
Across toxicity reasoning metrics, reasoning-aware training consistently improves reasoning quality, with particularly strong gains in causal coherence. This indicates that AOP-grounded supervision effectively aligns model-generated explanations with structured causal chains from MIEs to AOs. Hallucination scores show substantial improvement, with ToxReason-4B-GRPO outperforming all other models, indicating that reasoning-aware training effectively reduces unsupported statements. 

In contrast, biological fidelity improves only marginally, reflecting our training focus on AOP-defined mechanistic alignment instead of broad biological knowledge acquisition. Consequently, gains in overall reasoning quality are driven primarily by improved causal coherence and structural consistency, rather than by uniform improvements across all reasoning dimensions.

\subsection{Algorithmic Validation of LLM-Evaluator}
To assess whether LLM-as-a-Judge scores are supported by an objective signal, we examine their correlation with an algorithm-based measure of mechanistic reasoning alignment computed using the Needleman–Wunsch (NW) algorithm~\citep{needleman1970general}. NW is a dynamic programming method for global sequence alignment that preserves relative ordering while penalizing missing or extraneous steps. It has been adopted as a fine-grained metric to quantify alignment between generated and reference causal reasoning paths~\citep{nguyen2024direct}.

% Alignment highlighting
\providecommand{\alignA}{}
\providecommand{\alignB}{}
\providecommand{\alignC}{}
\providecommand{\alignD}{}
\providecommand{\alignE}{} % <- 추가

\renewcommand{\alignA}[1]{\begingroup\setlength{\fboxsep}{1pt}\colorbox{yellow!30}{#1}\endgroup}
\renewcommand{\alignB}[1]{\begingroup\setlength{\fboxsep}{1pt}\colorbox{cyan!25}{#1}\endgroup}
\renewcommand{\alignC}[1]{\begingroup\setlength{\fboxsep}{1pt}\colorbox{orange!35}{#1}\endgroup}
\renewcommand{\alignD}[1]{\begingroup\setlength{\fboxsep}{1pt}\colorbox{magenta!18}{#1}\endgroup}
\renewcommand{\alignE}[1]{\begingroup\setlength{\fboxsep}{1pt}\colorbox{green!25}{#1}\endgroup} % <- 신규
\begin{table*}[t]
\centering

\begin{tcolorbox}[
  width=\linewidth,
  colback=gray!3,
  colframe=black,
  title=Case Study: AOP Alignment \& Reasoning Quality (Qwen-4B-Inst. vs ToxReason-4B-GRPO),
  boxrule=0.8pt,
  arc=2pt,
  left=8pt,right=8pt,top=6pt,bottom=6pt
]

\begingroup
\setlength{\baselineskip}{0.93\baselineskip}
\setlist[itemize]{itemsep=-3pt, topsep=0pt, leftmargin=1.2em}
\setlist[enumerate]{itemsep=-2pt, topsep=0pt, leftmargin=1.6em}

\vspace{0.4em}
\textbf{SMILES:} \texttt{CC\#CC1(CCC2C1(CC(C3=C4CCC(=O)C=C4CCC23)C5=CC=C(C=C5)N(C)C)C)O}

\vspace{0.6em}
\textbf{Ground Truth AOP Context:}
\vspace{0.2em}

\begin{itemize}
  \item \textbf{AOP ID:} \texttt{Aop:318}
  \item \textbf{AOP Title:} Glucocorticoid Receptor activation leading to hepatic steatosis
  \item \textbf{AOP:}\\
  \hspace*{0.8em}\textit{\alignA{GR activation}} $\rightarrow$
  \textit{Decreased acyl-CoA dehydrogenase activity} $\rightarrow$
  \textit{\alignB{Reduced mitochondrial fatty} \alignB{acid $\beta$-oxidation}} $\rightarrow$
  \textit{\alignC{Triglyceride accumulation}} $\rightarrow$
  \textit{\alignD{Hepatic steatosis (liver toxicity)}}
\end{itemize}

\vspace{0.5em}
\noindent\rule{\linewidth}{0.4pt}
\vspace{0.6em}

% --- 2-column 비교 영역 ---
\noindent
\begin{tabularx}{\linewidth}{@{}>{\raggedright\arraybackslash}X
                            !{\vrule width 0.8pt}
                            >{\raggedright\arraybackslash}X@{}}

% ---------------- LEFT: Qwen ----------------
\textbf{Qwen-4B-Inst.}\par\vspace{0.3em}
\begin{itemize}
  \item \textbf{MIE:} \alignA{GR Activation}
  \item \textbf{Model Reasoning:}
\end{itemize}
\begin{enumerate}
  \item Activation of the \alignA{glucocorticoid receptor (GR)} leads to dysregulation of gene expression involved in metabolism and inflammation.
  \item This dysregulation results in increased hepatic lipid accumulation and oxidative stress, contributing to liver injury.
  \item Chronic liver injury progresses to hepatocyte necrosis and inflammation, leading to \alignD{liver toxicity}.
\end{enumerate}
\vspace{1.9em}
\begin{itemize}
  \item \textbf{Predicted Toxicity:} Liver toxicity
\end{itemize}

&

% ---------------- RIGHT: ToxReason ----------------
\textbf{ToxReason-4B-GRPO}\par\vspace{0.3em}
\begin{itemize}
  \item \textbf{MIE:} \alignA{GR Activation}
  \item \textbf{Model Reasoning:}
\end{itemize}
\begin{enumerate}
  \item \alignA{Glucocorticoid receptor (GR) activation} leads to \alignB{reduced hepatic fatty acid} \alignB{$\beta$-oxidation} in the liver.
  \item Reduced fatty acid $\beta$-oxidation leads to decreased fatty acid oxidation and increased fatty acid accumulation.
  \item Increased fatty acid accumulation leads to increased hepatic \alignC{triglyceride} \alignC{accumulation (steatosis)}.
  \item Increased \alignD{hepatic steatosis} leads to \alignD{liver toxicity}.
\end{enumerate}
\vspace{0.5em}
\begin{itemize}
  \item \textbf{Predicted Toxicity:} Liver toxicity
\end{itemize}

\end{tabularx}

\endgroup
\end{tcolorbox}

\captionsetup{skip=8pt, labelfont=bf}
\vspace{-10pt}
\caption{Case study comparing two LLMs on the same AOP grounding. The upper block provides the reference AOP context (Aop:318); the lower block contrasts model reasoning and predicted toxicity side-by-side. Highlighted segments denote aligned MIE/KE/AO. Additional reasoning evaluation results are reported in Table~\ref{tab:aop_case_study_combined}. Outputs from the other models and other sample of ToxReason-4B-GRPO are provided in Tables~\ref{tab:aop_case_study_example1} to \ref{tab:aop_case_study_toxreason_grpo}.}
\label{tab:aop_case_study_qwen_vs_toxreason}
\vspace{-8pt}
\end{table*}

As mechanistic toxicity reasoning follows an ordered causal progression from MIEs to AOs, NW provides a structure-aware measure of causal consistency by aligning semantically encoded reasoning steps with reference AOP steps to compute a global alignment score that accounts for semantic similarity and causal ordering (see Appendix~\ref{sec:appendix_GRPO} \texttt{tox\_align\_score} and Algorithm~\ref{alg:alignment-score}). 

Table~\ref{tab:nw_alignment} reports that the overall LLM-as-a-Judge reasoning score shows the strongest correlation with NW-alignment Score (Pearson $r=0.703$, Spearman $\rho=0.837$), indicating that holistic judge evaluations reflect adherence to the intended mechanistic structure, thereby supporting the reliability of LLM-as-a-Judge as an evaluation signal.

\subsection{Case Study}
As shown in Table~\ref{tab:aop_case_study_qwen_vs_toxreason}, we analyze the reasoning outputs of a base model and the ToxReason-4B-GRPO under the same AOP grounding. The base model correctly identifies the molecular initiating event and predicts liver toxicity, but its explanation remains generic and loosely structured, omitting several intermediate biological events specified in the reference AOP. 
In contrast, ToxReason-4B-GRPO generates a step-by-step mechanistic explanation that closely follows the reference AOP causal chain, explicitly tracing intermediate processes from glucocorticoid receptor activation through reduced hepatic fatty acid $\beta$-oxidation and triglyceride accumulation to hepatic steatosis. This comparison shows that reasoning-aware training improves the structural fidelity of mechanistic explanations.
\section{Conclusion}
In this work, we introduced ToxReason, a benchmark designed to evaluate mechanistic toxicity reasoning grounded in AOPs. ToxReason moves beyond outcome-level toxicity prediction and enables systematic assessment of causal reasoning from molecular initiating events to organ-level adverse outcomes. Furthermore, we established a toxicity-reasoning-aware model to advance the use of LLMs in toxicology. These contributions facilitate a deeper mechanistic understanding while enabling more reliable and interpretable toxicity assessments. Ultimately, we envision that such mechanistically grounded reasoning frameworks could assist regulatory decision-making processes in drug safety evaluation.

\clearpage
\section*{Limitations}
Despite its contributions, this work has several limitations. First, ToxReason is constrained by the coverage and granularity of AOP-Wiki, and our benchmark currently focuses on only liver, heart, and kidney toxicities, limiting generalization to other organs or less-characterized mechanisms. In addition, the set of MIEs considered in ToxReason is restricted to those defined within the selected AOPs, resulting in a limited and domain-specific MIE space. Second, the reference mechanistic pathways used for evaluation represent canonical causal structures and may not fully capture the diversity or context-dependence of real-world toxicity mechanisms.
Furthermore, MIE predictions for query compounds are inferred by referencing MIE annotations from structurally similar molecules rather than being derived solely from molecular structure. This design choice was intentionally adopted to control for variability in molecular representation learning and to focus the evaluation on the model’s ability to perform mechanistic toxicity reasoning, rather than its capacity to independently infer molecular interactions. Finally, while LLM-as-a-Judge evaluation enables scalable assessment of reasoning quality, it remains inherently subjective. Although we mitigate this by complementing it with an algorithm-based alignment metric, judge-based scores should be interpreted as relative rather than absolute measures.

\section*{Ethics Statement}

This research utilizes strictly publicly available datasets, including AOP-Wiki, ChEMBL, and CTD, and does not involve human subjects, personal data, or animal experimentation. All data were accessed and analyzed in compliance with their original licensing terms and intended research objectives. ToxReason is developed solely for research and evaluation purposes; consequently, model outputs must not be interpreted as standalone evidence for clinical or regulatory decision-making. We explicitly acknowledge the potential risks associated with over-reliance on automated systems in safety-critical domains such as toxicology. Accordingly, model-generated insights should be used with caution and must be integrated with expert judgment and empirical validation.

\section*{Acknowledgments}
This research was supported by (1) the National Research Foundation of Korea (NRF-2023R1A2C3004176), (2) the Ministry of Health \& Welfare, Republic of Korea (HR20C002103), (3) ICT Creative Consilience Program through the Institute of Information \& Communications Technology Planning \& Evaluation (IITP) grant funded by the Korea government (MSIT) (IITP-2025 RS-2020-II201819), (4) the National Research Foundation of Korea(NRF) grant funded by the Korea governmant(MSIT and MOE) (No. RS-2025-16652968), (5) the Seoul National University Hospital with support from the Ministry of Science and ICT (RS- 2023-00262002) and (6) the Korea Bio Data Station(K-BDS) with computing resources including technical support.

\bibliographystyle{acl_natbib}
% \bibliography{9_reference}

\clearpage
\section*{Appendix}

\appendix
\renewcommand{\thetable}{\Alph{table}}
\setcounter{table}{0}
\section{Details in ToxReason Benchmark}
\label{sec:appendix_details}
\paragraph{Selected AOPs} Table \ref{tab:aop_selection} summarizes the curated set of adverse outcome pathways (AOPs) used in ToxReason, including their associated adverse outcomes (AOs) and molecular initiating events (MIEs), with activation and inhibition directions explicitly annotated.
\FloatBarrier
\begin{table}[t]
\centering
\setlength{\tabcolsep}{6pt}
\renewcommand{\arraystretch}{1.35}

\resizebox{\columnwidth}{!}{%
\begin{tabular}{c c l c}
\noalign{\hrule height 1.5pt}
\textbf{Toxicity} & \textbf{AO (Disease)} & \textbf{AOP (ID)} & \textbf{MIE} \\
\hline

\multirow{14}{*}{\makecell{\Large Liver\\\Large Toxicity}}
& \multirow{3}{*}{\makecell{\Large Liver\\\Large Cancer}}
& AOP:37  & PPAR$\alpha$ $\uparrow$ \\
&  & AOP:41  & AHR $\uparrow$ \\
&  & AOP:220 & CYP2E1 $\uparrow$ \\[8pt]

& \multirow{5}{*}{\makecell{\Large Liver\\\Large Fibrosis}}
& AOP:130 & PLA2 $\uparrow$ \\
&  & AOP:494 & AHR $\uparrow$ \\
&  & AOP:34  & LXR $\uparrow$ \\
&  & AOP:36  & PPAR$\alpha/\beta/\gamma$ $\downarrow$ \\
&  & AOP:57  & AHR $\uparrow$ \\[8pt]

& \multirow{6}{*}{\makecell{\Large Liver\\\Large Steatosis}}
& AOP:58  & LXR $\uparrow$, PPAR$\alpha$ $\downarrow$ \\
&  & AOP:60  & PXR $\uparrow$ \\
&  & AOP:61  & NR1H4 $\uparrow$, NRF2 $\uparrow$ \\
&  & AOP:318 & GR $\uparrow$ \\
&  & AOP:517 & PXR $\uparrow$ \\
&  & AOP:518 & LXR $\uparrow$ \\

\hline

\multirow{7}{*}{\Large Cardiotoxicity}
& \multirow{4}{*}{\makecell{\Large Heart\\\Large Failure}}
& AOP:553 & SCN5A $\downarrow$ \\
&  & AOP:555 & KCNH2 $\downarrow$ \\
&  & AOP:558 & PDE3/4 $\downarrow$ \\
&  & AOP:554 & ADRB2 $\uparrow$ \\[8pt]

& \multirow{3}{*}{\makecell{\Large Cardiac\\\Large Arrhythmia}}
& AOP:559 & AChE $\downarrow$ \\
&  & AOP:138 & OAT1 $\downarrow$ \\
&  & AOP:177 & COX1 $\downarrow$ \\

\hline

\multirow{4}{*}{\makecell{\Large Kidney\\\Large Toxicity}}
& \multirow{2}{*}{\makecell{\Large Kidney\\\Large Failure}}
& AOP:138 & OAT1 $\downarrow$ \\
&  & AOP:177 & COX1 $\downarrow$ \\[8pt]

& \makecell{\Large Chronic\\\Large Kidney Disease}
& AOP:384 & ACE $\downarrow$, AT1R $\uparrow$ \\

\noalign{\hrule height 1.5pt}
\end{tabular}
}

\caption{Selected adverse outcome pathways (AOPs) with associated adverse outcomes (AOs) and molecular initiating events (MIEs). ($\uparrow$: Activation, $\downarrow$: Inhibition)}
\label{tab:aop_selection}
\end{table}

\begin{table}[t]
\centering
\setlength{\tabcolsep}{6pt}
\renewcommand{\arraystretch}{1.5}

\resizebox{\columnwidth}{!}{%
{\large
\begin{tabular}{cccc}
\noalign{\hrule height 2pt}
\textbf{Toxicity Label}
& \makecell[c]{\textbf{MIE-matched}\\\textbf{Trainset}}
& \makecell[c]{\textbf{MIE-AO-matched}\\\textbf{Trainset}}
& \makecell[c]{\textbf{Testset}} \\
\hline
Liver                 & 10237 & 1635 & 63  \\
Cardio                & 18543 & 906  & 75  \\
Kidney                & 132   & 72   & 39  \\
Liver, Cardio         & 91    & 764  & 4   \\
Liver, Kidney         & 0     & 285  & 5   \\
Cardio, Kidney        & 1197  & 296  & 7   \\
Liver, Cardio, Kidney & 12    & 1279 & 0   \\
\textbf{Total}        & \textbf{30212} & \textbf{5237} & \textbf{193} \\
\noalign{\hrule height 2pt}
\end{tabular}
}
}

\caption{Sample distribution by toxicity category for MIE-matched and MIE-AO-matched training sets and the test set.}
\label{tab:statistics}
\end{table}

\paragraph{Data Statistics} Table \ref{tab:statistics} reports the distribution of samples across toxicity labels in the MIE-matched and MIE–AO-matched training sets, as well as the test set, providing an overview of dataset composition and label coverage.

\section{Training Details}
\label{sec:appendix_training_details}
\subsection{In-Context Learning} For in-context learning, we augment the prompt with a small number of demonstration examples sampled from a MIE-AO-matched trainset. For each query molecule, examples are randomly selected according to the specified shot number and follow the same input structure as the original task, including the corresponding output formatted in JSON. We evaluate in-context learning under 1, 2 and 4-shot configurations to examine the effect of increasing contextual supervision at inference time.

\subsection{Supervised Fine-Tuning}
For supervised fine-tuning, we used Qwen3-4B-Instruct as the base model and performed instruction tuning on the MIE-AO-matched training set. Each training instance follows the task formulation of ToxReason, consisting of structured inputs and corresponding JSON-formatted outputs that encode mechanistic toxicity reasoning.
Parameter-efficient fine-tuning was conducted using LoRA, with a rank of 8, a scaling factor ($\alpha$) of 32, and a dropout rate of 0.1.
The model was trained with a learning rate of $2 \times 10^{-5}$ using a cosine learning rate scheduler with a warmup ratio of 0.1, for a total of 3 epochs.
We used a per-device batch size of 8 and applied gradient accumulation with 8 steps.
All experiments were conducted using 8 NVIDIA A100 GPUs.

\subsection{Reinforcement Learning}
\label{sec:appendix_GRPO}
Reinforcement learning was conducted in two stages to progressively enhance the model’s mechanistic reasoning capability.
\paragraph{Stage 1}
In the first stage, the model was initialized by performing supervised fine-tuning on the MIE-matched training set. 
We used Qwen3-4B-Instruct as the base model and applied LoRA for parameter-efficient fine-tuning, with rank $r{=}32$, scaling factor $\alpha{=}64$, and dropout rate of 0.05.
The model was trained with a learning rate of $5\times10^{-5}$ using a cosine learning rate scheduler with a warmup ratio of 0.03, for 5 epochs.
We used a per-device batch size of 8 with gradient accumulation over 8 steps.

\paragraph{Stage 2}
In the second stage, we further trained the model using Group Relative Policy Optimization (GRPO) on the MIE-AO-matched training set.
Training was performed with a per-device batch size of 8 and gradient accumulation over 4 steps.
We used a learning rate of $1.0\times10^{-6}$ with a cosine scheduler with a minimum learning rate, and trained the model for a maximum of 1000 steps.
For each input, the model generated 8 candidate responses, using a sampling temperature of 0.7.

During this stage, we employed three reward functions to guide the learning process, which are described in detail below.
\begin{itemize}
  \item \textbf{\texttt{tox\_format}} (format compliance).
This reward encourages the model to produce a valid JSON output that strictly follows the predefined schema.
Specifically, it rewards successful JSON parsing, exact compliance with the required top-level fields, and a one-to-one correspondence between predicted MIEs and reasoning blocks.
Additional constraints are imposed to ensure well-formed reasoning blocks, consistent label usage, and the absence of malformed or multi-line fields.
The final score is normalized to the range $[0,1]$.

\item \textbf{\texttt{tox\_mie\_pred}} (MIE prediction accuracy).
This reward measures the accuracy of predicted MIEs by comparing them with the reference answer.
Target-level agreement is quantified using the Jaccard similarity between the sets of predicted and reference MIEs, while directional consistency is assessed by checking whether activation or inhibition labels match for overlapping targets.
To discourage over-prediction, the reward applies a penalty when the model predicts more MIEs than present in the reference.
The final score reflects both target overlap and direction correctness and is normalized to the range $[0,1]$.

\item \textbf{\texttt{tox\_align\_score}} (alignment score).
This reward evaluates the causal consistency of the generated reasoning by measuring how well it aligns with the Answer AOP in an order-preserving manner.
Each reasoning step is first embedded using the \texttt{all-MiniLM-L6-v2} sentence embedding model, and step-level similarities are computed based on cosine similarity.
The overall alignment score is then computed using the Needleman-Wunsch(NW) global sequence alignment algorithm, which preserves the relative order of causal steps while penalizing missing or extraneous steps.
We use an acceptance threshold of 0.20 for candidate matches, a similarity threshold of 0.50, a gap penalty of $-0.30$, and a low-similarity penalty of $-0.15$.
The resulting alignment score is normalized and clipped to the range $[0,1]$. The same alignment score formulation is also applied when evaluating the model's reasoning outputs.
\end{itemize}
The three reward components were assigned equal weights during optimization.
Reinforcement learning experiments were carried out using 8 NVIDIA A100 GPUs.
\begin{algorithm}[t]
\caption{NW-Alignment Score Computation}
\label{alg:alignment-score}
\begin{algorithmic}[1]

\Statex \textbf{Input:}
\Statex \parbox[t]{0.9\linewidth}{
Answer AOP path $A = (a_1,\dots,a_M)$;\\
Pred. reasoning path $P = (p_1,\dots,p_N)$;\\
Step similarity function $\mathrm{sim}(\cdot,\cdot)$;\\
Gap penalty $\delta < 0$
}

\Statex \textbf{Output:}
Alignment score $S \in [0,1]$

\State Initialize score matrix $D \in \mathbb{R}^{(M+1)\times(N+1)}$
\State $D[0,0] \gets 0$

\For{$i = 1$ to $M$}
    \State $D[i,0] \gets D[i-1,0] + \delta$
\EndFor

\For{$j = 1$ to $N$}
    \State $D[0,j] \gets D[0,j-1] + \delta$
\EndFor

\For{$i = 1$ to $M$}
    \For{$j = 1$ to $N$}
        \State $match \gets D[i-1,j-1] + \mathrm{sim}(a_i, p_j)$
        \State $skip\_answer \gets D[i-1,j] + \delta$
        \State $skip\_pred \gets D[i,j-1] + \delta$
        \State $D[i,j] \gets \max(match, skip\_answer, skip\_pred)$
    \EndFor
\EndFor

\State $raw \gets D[M,N]$
\State $max \gets \min(M,N)$
\State $min \gets (M+N)\cdot \delta$
\State $S \gets \dfrac{raw - min}{max - min}$

\State \Return $S$

\end{algorithmic}
\end{algorithm}

\FloatBarrier

\begin{table*}[t]
\centering

\begin{tcolorbox}[
  width=\linewidth,
  colback=blue!3,
  colframe=blue!70!black,
  colbacktitle=blue!80!black,
  coltitle=white,
  title=LLM-as-a-Judge \textbf{SYSTEM PROMPT},
  boxrule=0.9pt,
  arc=3pt,
  left=10pt,
  right=10pt,
  top=6pt,
  bottom=8pt
]

\footnotesize
\setlength{\baselineskip}{0.93\baselineskip}

You are a toxicology expert and AOP-based mechanistic reasoning evaluator.\\
Your task is to assess the quality of an AI assistant's toxicity reasoning based on:
\begin{itemize}[leftmargin=1.2em, itemsep=2pt, topsep=2pt]
  \item its final predicted toxicities,
  \item its step-by-step mechanistic reasoning,
  \item and the provided ground-truth AOP and toxicity labels.
\end{itemize}

\vspace{0.4em}
Evaluate the assistant’s response using the following criteria, each on a scale from 1 to 10
(higher scores are better):

\begin{enumerate}[leftmargin=1.2em, itemsep=4pt, topsep=2pt]
  \item Hallucination\_Avoidance --- The degree to which the model avoids inventing unsupported facts.
  A high score means little to no hallucination and strong grounding in the provided AOP and inputs.

  \item Causal\_Coherence --- Logical consistency of the mechanistic chain.
  Each step should follow causally from the previous one
  (\textit{MIE} $\rightarrow$ \textit{KE} $\rightarrow$ \textit{AO} $\rightarrow$ \textit{Organ Toxicity})
  without unjustified jumps, contradictions, or reversed order.

  \item Biological\_Fidelity --- Biological validity of the mechanism.
  Uses correct terminology, accurate MIE/KE/AO relationships, and reflects realistic heart/liver/kidney
  toxicology and physiology.

  \item Overall --- An overall quality score summarizing the four criteria above.
\end{enumerate}

\vspace{0.5em}
\textbf{Output Format Requirement:}\\
You must output a single valid JSON object with the following structure:

\vspace{0.35em}
\begin{verbatim}
{
  "Hallucination_Avoidance": <number from 1 to 10>,
  "Causal_Coherence": <number from 1 to 10>,
  "Biological_Fidelity": <number from 1 to 10>,
  "Overall": <number from 1 to 10>,
  "Explanation": "<short textual explanation (3-6 sentences)>"
}
\end{verbatim}

\vspace{0.4em}
\textbf{Hard Constraints:}
\begin{itemize}[leftmargin=1.2em, itemsep=2pt, topsep=2pt]
  \item Output only the JSON object.
  \item Do not include any additional text, comments, or markdown.
  \item The JSON must be syntactically valid.
\end{itemize}

\end{tcolorbox}

\captionsetup{skip=8pt, labelfont=bf}
\caption{LLM-as-a-judge system prompt used for evaluating mechanistic toxicity reasoning quality.}
\label{tab:llm_as_judge_system_prompt}
\end{table*}
\begin{table*}[t]
\centering

\begin{tcolorbox}[
  width=\linewidth,
  colback=blue!3,
  colframe=blue!70!black,
  colbacktitle=blue!80!black,
  coltitle=white,
  title=LLM-as-a-Judge \textbf{USER PROMPT},
  boxrule=0.9pt,
  arc=3pt,
  left=10pt,
  right=10pt,
  top=6pt,
  bottom=8pt,
  enhanced,
  sharp corners
]

\footnotesize
\setlength{\baselineskip}{0.93\baselineskip}

\textbf{[Ground-truth Information]}\\[0.3em]

Correct\_AOP:\\
\texttt{\{\{AOP\_Context\}\}}

\vspace{0.4em}
Correct\_Toxicities:\\
\texttt{\{\{Answer\_TOXICITIES\}\}}

\vspace{0.6em}
\textbf{[Assistant Response to Evaluate]}\\[0.3em]

\texttt{\{\{Assistant\_Response\}\}}

\vspace{0.4em}
Please evaluate the assistant’s reasoning according to the criteria defined in the system prompt
and return only the JSON object with your scores and explanation.

\end{tcolorbox}

\captionsetup{skip=8pt, labelfont=bf}
\caption{LLM-as-a-judge user prompt for evaluating toxicity reasoning against ground-truth AOP annotations.}
\label{tab:llm_as_judge_user_prompt}

\end{table*}

\begin{table*}[t]
\centering
\begin{tcolorbox}[
  width=\linewidth,
  colback=blue!3,                % ← 배경만 파란색
  colframe=blue!70!black,        % ← 테두리 파란색
  colbacktitle=blue!80!black,    % ← 타이틀 바
  coltitle=white,
  title=LLM Inference \textbf{SYSTEM PROMPT},
  boxrule=0.9pt,
  arc=3pt,
  left=10pt,
  right=10pt,
  top=6pt,
  bottom=8pt
]

\footnotesize

\begingroup
\setlength{\baselineskip}{0.93\baselineskip}
\vspace{0.7em}

You are an expert mechanistic toxicology reasoning model.\\
\\
Given:\\
- A query molecule (SMILES)\\
- Experimental observations from structurally similar molecules, including
  Activation Examples and Inhibition Examples with similarity scores,\\

your job is to infer how the query molecule interacts with biological targets (MIEs) and then explain how these inferred MIEs could mechanistically lead to organ toxicity.\\
\\
Follow these rules carefully:\\
\\
1. When interpreting reference evidence:\\
    - Use ONLY Activation Examples to infer activation.\\
    - Use ONLY Inhibition Examples to infer inhibition.\\
    - Never infer inhibition from “non-active”.\\
    - Never infer activation from “non-inhibit”.\\
    - Give more weight to examples with higher similarity scores.\\
   - Base all conclusions on structural similarity + target evidence.\\

2.  For EACH inferred MIE, produce mechanistic reasoning describing how it can lead to organ toxicity:\\
   - Use “Step 1, Step 2, Step 3...” format.\\
   - Each step $\leq$ 2 sentences.\\
   - Steps must follow: \\
       MIE → Key Events (KEs) → Adverse Outcome (AO) → organ toxicity.\\
\\
3. Only consider the following toxicity types:\\
   - “cardiotoxicity”\\
   - “liver toxicity”\\
   - “kidney toxicity”\\
   Choose exactly ONE organ toxicity per MIE.
\begin{verbatim}
{
  "MIE_Prediction": {
    "Target1": "Activation or Inhibition",
    "Target2": "Activation or Inhibition",
    ...
  },
  "Toxicity_Reasoning": [
    {
      "MIE": "",
      "Reasoning_Steps": [
        "Step 1: ...",
        "Step 2: ...",
        "Step 3: ..."
      ],
      "Toxicity": ""
    }
  ],
  "Overall_Assessment": {
    "Summary": "Concise narrative summary of why toxicity occurs.",
    "Predicted_Toxicities": ["...", "..."]
  }
}
\end{verbatim}

HARD CONSTRAINTS:\\
- Output ONLY the JSON.\\
- Activation/Inhibition labels must be exactly: "Activation", "Inhibition".\\
- Toxicity must be one of the three organ toxicities listed.\\
- No extra text outside the JSON.

\endgroup
\end{tcolorbox}

\captionsetup{skip=8pt, labelfont=bf}
\caption{LLM inference SYSTEM PROMPT}
\label{tab:llm_inference_system_prompt}
\end{table*}

\begin{table*}[t]
\centering
\begin{minipage}{\linewidth}
\begin{tcolorbox}[
  enhanced,
  width=\linewidth,
  colback=blue!3,
  colframe=blue!70!black,
  colbacktitle=blue!80!black,
  coltitle=white,
  title=LLM Inference \textbf{USER PROMPT},
  boxrule=0.9pt,
  arc=3pt,
  left=10pt,
  right=10pt,
  top=6pt,
  bottom=8pt
]

\footnotesize
\setlength{\baselineskip}{0.93\baselineskip}

Query Molecule (SMILES):\\
\texttt{\{\{test\_smiles\}\}}

\vspace{0.4em}
Below is experimental evidence from structurally similar molecules.
Each example shows a similarity score and activity result on specific biological targets.

\vspace{0.4em}
Reference Evidence:\\
\texttt{\{\{refer\_context\}\}}

\vspace{0.6em}
Using the evidence above:
\begin{itemize}[leftmargin=1.2em, itemsep=2pt, topsep=2pt]
  \item Interpret the activation/inhibition patterns based on similarity.
  \item Infer the most likely molecular initiating events (MIEs) of the query molecule.
  \item Produce mechanistic toxicity reasoning showing how those MIEs could lead to
  organ-level toxicity through the sequence:\\
  \hspace*{1.2em}MIE $\rightarrow$ Key Events $\rightarrow$ Adverse Outcome $\rightarrow$ Organ Toxicity.
\end{itemize}

\vspace{0.3em}
Return ONLY the final JSON in the required format.

\end{tcolorbox}
\end{minipage}

\captionsetup{skip=8pt, labelfont=bf}
\caption{LLM inference user prompt used for mechanistic toxicity reasoning.}
\label{tab:llm_inference_user_prompt}
\end{table*}
\FloatBarrier

\FloatBarrier
% \section{Case Study}
% \label{sec:appendix_case_study}

\begin{table*}[t]
\centering

% --- scale the whole tcolorbox to fit within one page height ---
\resizebox{!}{0.86\textheight}{%
\begin{tcolorbox}[
  width=\linewidth,
  colback=gray!3,
  colframe=black,
  title=Case Study: AOP Alignment \& Reasoning Quality (Qwen-4B-Inst. vs ToxReason-4B-GRPO),
  boxrule=0.8pt,
  arc=2pt,
  left=8pt,right=8pt,top=6pt,bottom=6pt
]

\begingroup
\setlength{\baselineskip}{0.93\baselineskip}
\setlist[itemize]{itemsep=-3pt, topsep=0pt, leftmargin=1.2em}
\setlist[enumerate]{itemsep=-2pt, topsep=0pt, leftmargin=1.6em}

\vspace{0.4em}
\textbf{SMILES:} \texttt{CC\#CC1(CCC2C1(CC(C3=C4CCC(=O)C=C4CCC23)C5=CC=C(C=C5)N(C)C)C)O}

\vspace{0.6em}
\textbf{Ground Truth AOP Context:}
\vspace{0.2em}

\begin{itemize}
  \item \textbf{AOP ID:} \texttt{Aop:318}
  \item \textbf{AOP Title:} Glucocorticoid Receptor activation leading to hepatic steatosis
  \item \textbf{AOP:}\\
  \hspace*{0.8em}\textit{\alignA{GR activation}} $\rightarrow$
  \textit{Decreased acyl-CoA dehydrogenase activity} $\rightarrow$
  \textit{\alignB{Reduced mitochondrial fatty} \alignB{acid $\beta$-oxidation}} $\rightarrow$
  \textit{\alignC{Triglyceride accumulation}} $\rightarrow$
  \textit{\alignD{Hepatic steatosis (liver toxicity)}}
\end{itemize}

\vspace{0.35em}
\noindent\rule{\linewidth}{0.4pt}
\vspace{0.35em}

% ===================== (A) LLM RESPONSES (highlighted) =====================
\noindent
\begin{tabularx}{\linewidth}{@{}>{\raggedright\arraybackslash}X
                            !{\vrule width 0.8pt}
                            >{\raggedright\arraybackslash}X@{}}

% -------- LEFT: Qwen response --------
\textbf{Qwen-4B-Inst.}\par\vspace{0.2em}
\begin{itemize}
  \item \textbf{MIE:} \alignA{GR Activation}
  \item \textbf{Model Reasoning:}
\end{itemize}
\begin{enumerate}
  \item Activation of the \alignA{glucocorticoid receptor (GR)} leads to dysregulation of gene expression involved in metabolism and inflammation.
  \item This dysregulation results in increased hepatic lipid accumulation and oxidative stress, contributing to liver injury.
  \item Chronic liver injury progresses to hepatocyte necrosis and inflammation, leading to \alignD{liver toxicity}.
\end{enumerate}
\vspace{1.7em}
\begin{itemize}
  \item \textbf{Predicted Toxicity:} Liver toxicity
\end{itemize}

&

% -------- RIGHT: ToxReason response --------
\textbf{ToxReason-4B-GRPO}\par\vspace{0.2em}
\begin{itemize}
  \item \textbf{MIE:} \alignA{GR activation}
  \item \textbf{Model Reasoning:}
\end{itemize}
\begin{enumerate}
  \item \alignA{Glucocorticoid receptor (GR) activation} leads to \alignB{reduced hepatic fatty acid} \alignB{$\beta$-oxidation} in the liver.
  \item Reduced fatty acid $\beta$-oxidation leads to decreased fatty acid oxidation and increased fatty acid accumulation.
  \item Increased fatty acid accumulation leads to increased hepatic \alignC{triglyceride} \alignC{accumulation (steatosis)}.
  \item Increased \alignD{hepatic steatosis} leads to \alignD{liver toxicity}.
\end{enumerate}
\vspace{0.2em}
\begin{itemize}
  \item \textbf{Predicted Toxicity:} Liver toxicity
\end{itemize}

\end{tabularx}

% ===================== divider between response and judge =====================
\vspace{0.15em}
\noindent\rule{\linewidth}{0.4pt}
\vspace{0.15em}

% ===================== (B) LLM-as-a-Judge OUTPUT (NO highlight) =====================
\noindent\textbf{LLM-as-a-Judge Evaluation:}\par\vspace{0.15em}

\noindent
\begin{tabularx}{\linewidth}{@{}>{\raggedright\arraybackslash}X
                            !{\vrule width 0.8pt}
                            >{\raggedright\arraybackslash}X@{}}

% -------- LEFT: Qwen judge --------
\textbf{Qwen-4B-Inst.}\par\vspace{0.2em}
\begin{itemize}
  \item \textbf{Judge Scores:}
  \begin{itemize}
    \item Hallucination Avoidance: \textbf{4}
    \item Causal Coherence: \textbf{5}
    \item Biological Fidelity: \textbf{5}
    \item Overall: \textbf{4}
  \end{itemize}
  \item \textbf{Judge Summary:}
\end{itemize}
\begin{enumerate}
  \item Correctly identifies glucocorticoid receptor (GR) activation as the initiating molecular event.
  \item Mentions general lipid accumulation but omits critical intermediate steps such as decreased acyl-CoA dehydrogenases and impaired mitochondrial $\beta$-oxidation.
  \item Introduces unsupported mechanisms (oxidative stress, inflammation, hepatocyte necrosis) not specified in the ground-truth AOP.
  % \item Hallucinates an additional molecular initiating event (KCNH2 inhibition) and predicts cardiotoxicity, which is not present in the provided AOP.
\end{enumerate}

&

% -------- RIGHT: ToxReason judge --------
\textbf{ToxReason-4B-GRPO}\par\vspace{0.2em}
\begin{itemize}
  \item \textbf{Judge Scores:}
  \begin{itemize}
    \item Hallucination Avoidance: \textbf{8}
    \item Causal Coherence: \textbf{9}
    \item Biological Fidelity: \textbf{8}
    \item Overall: \textbf{8}
  \end{itemize}
  \item \textbf{Judge Summary:}
\end{itemize}
\begin{enumerate}
  \item Glucocorticoid receptor activation leads to reduced mitochondrial fatty acid $\beta$-oxidation.
  \item Reduced $\beta$-oxidation results in increased fatty acid availability in hepatocytes.
  \item Excess fatty acids are stored as triglycerides, causing hepatic steatosis.
  \item Hepatic steatosis is directly linked to liver toxicity.
\end{enumerate}

\end{tabularx}

\endgroup
\end{tcolorbox}%
} % end resizebox

\captionsetup{skip=6pt, labelfont=bf}
\caption{Case study combining model-generated AOP reasoning (highlighted for alignment) and LLM-as-a-Judge evaluation (no highlighting). The upper block shows the reference AOP context and each model's reasoning; the lower block reports judge scores and diagnostic summaries for the same responses.}
\label{tab:aop_case_study_combined}
\end{table*}

% Alignment highlighting
\providecommand{\alignA}{}
\providecommand{\alignB}{}
\providecommand{\alignC}{}
\providecommand{\alignD}{}
\providecommand{\alignE}{} % <- 추가

\renewcommand{\alignA}[1]{\begingroup\setlength{\fboxsep}{1pt}\colorbox{yellow!30}{#1}\endgroup}
\renewcommand{\alignB}[1]{\begingroup\setlength{\fboxsep}{1pt}\colorbox{cyan!25}{#1}\endgroup}
\renewcommand{\alignC}[1]{\begingroup\setlength{\fboxsep}{1pt}\colorbox{orange!35}{#1}\endgroup}
\renewcommand{\alignD}[1]{\begingroup\setlength{\fboxsep}{1pt}\colorbox{magenta!18}{#1}\endgroup}
\renewcommand{\alignE}[1]{\begingroup\setlength{\fboxsep}{1pt}\colorbox{green!25}{#1}\endgroup} % <- 신규
\begin{table*}[t]
\centering
\begin{tcolorbox}[
  width=\linewidth,
  colback=gray!3,
  colframe=black,
  title=Case Study: AOP Alignment \& Reasoning Quality (GPT-5)
]
Below is an example instance illustrating strong alignment between the \textbf{Ground Truth AOP context (Aop:559)} and the \textbf{GPT-5} reasoning path, demonstrating accurate reproduction of an AChE-mediated cardiotoxicity pathway.

\begingroup
\setlength{\baselineskip}{0.93\baselineskip}
\vspace{0.7em}

\textbf{Sample Metadata.}
\vspace{0.4em}
\begin{itemize}[itemsep=-3pt, topsep=0pt, leftmargin=1.2em]
  \item \textbf{Index:} 98
  \item \textbf{SMILES:} \texttt{CC12CCN(C1N(C3=C2C=C(C=C3)OC(=O)NC)C)C}
  \item \textbf{AOP ID:} \texttt{Aop:559}
\end{itemize}

\vspace{0.7em}
\textbf{Ground Truth AOP Context:}
\vspace{0.4em}
\begin{itemize}[itemsep=-3pt, topsep=0pt, leftmargin=1.2em]
  \item \textbf{AOP Title:} Inhibition of acetylcholinesterase (AChE) leading to arrhythmias
  \item \textbf{AOP:}
  \hspace*{0.8em}\textit{\alignA{Acetylcholinesterase (AChE) Inhibition}} $\rightarrow$
  \textit{\alignB{Increased Muscarinic Acetylcholine}} \\$\rightarrow$
  \textit{\alignB{Receptors} Altered Action Potential} $\rightarrow$
  \textit{\alignC{Increased delay in heart electrical conduction}} $\rightarrow$
  \textit{\alignD{Occurrence of cardiac arrhythmia}}
\end{itemize}

\vspace{0.7em}
\textbf{GPT-5 Reasoning Result}
\vspace{0.4em}
\begin{itemize}[itemsep=-3pt, topsep=0pt, leftmargin=1.2em]
  \item \textbf{MIE:} \alignA{AChE Inhibition}
  \item \textbf{GPT-5 Reasoning:}
\end{itemize}

\begin{enumerate}[itemsep=-2pt, topsep=0pt, leftmargin=2.0em]
  \item \alignA{Inhibition of acetylcholinesterase} increases synaptic acetylcholine in cardiac parasympathetic pathways.
  \item Excess acetylcholine overstimulates \alignB{M2 muscarinic receptors},
        leading to slowed cardiac electrical conduction and increased vagal tone.
  \item Resulting electrical disturbances progress to \alignD{arrhythmias} and
        impaired cardiac output, culminating in cardiotoxicity.
\end{enumerate}

\begin{itemize}[itemsep=-3pt, topsep=0pt, leftmargin=1.2em]
  \item \textbf{Predicted Toxicity:} cardiotoxicity
\end{itemize}

\vspace{0.4em}
\textbf{GPT-5 Final Predicted Toxicity: cardiotoxicity}
\vspace{0.4em}

\endgroup
\end{tcolorbox}

\captionsetup{skip=8pt, labelfont=bf}
\caption{Case study example illustrating strong alignment between an AChE-mediated Gold AOP (Aop:559) and the GPT-5 reasoning path. Highlighted segments denote aligned molecular initiating events, key events, and adverse outcomes along the AOP cascade.}
\label{tab:aop_case_study_example1}
\end{table*}
\begin{table*}[t]
\centering
\begin{tcolorbox}[
  width=\linewidth,
  colback=gray!3,
  colframe=black,
  title=Case Study: AOP Alignment \& Reasoning Quality (GPT-5.1)
]
Below is an example instance illustrating alignment between the \textbf{Gold AOP context} and the \textbf{GPT-5.1} reasoning path for a renal toxicity case driven by COX1 inhibition.

\begingroup
\setlength{\baselineskip}{0.93\baselineskip}
\vspace{0.7em}

\textbf{Sample Metadata.}
\vspace{0.4em}
\begin{itemize}[itemsep=-3pt, topsep=0pt, leftmargin=1.2em]
  \item \textbf{Index:} 70
  \item \textbf{SMILES:} \texttt{CC(C)C(=O)NC1=CC(=C(C=C1)[N+](=O)[O-])C(F)(F)F}
\end{itemize}

\vspace{0.7em}
\textbf{Ground Truth AOP Context:}
\vspace{0.4em}
\begin{itemize}[itemsep=-3pt, topsep=0pt, leftmargin=1.2em]
  \item \textbf{AOP ID:} \texttt{Aop:177}
  \item \textbf{AOP Title:} Cyclooxygenase 1 (COX1) inhibition leading to renal failure and mortality
  \item \textbf{AOP:} \\
  \hspace*{0.8em}\textit{\alignA{Inhibition of Cyclooxygenase 1 (COX1)}} $\rightarrow$
  \textit{\alignB{Decreased prostaglandin levels}} $\rightarrow$
  \textit{\alignC{Renal ischemia and oxidative stress}} $\rightarrow$
  \textit{\alignD{Renal proximal tubular necrosis}} $\rightarrow$
  \textit{\alignD{Renal failure and increased mortality}}
\end{itemize}

\vspace{0.7em}
\textbf{GPT-5.1 Reasoning Response}
\vspace{0.4em}
\begin{itemize}[itemsep=-3pt, topsep=0pt, leftmargin=1.2em]
  \item \textbf{MIE:} \alignA{COX1 Inhibition}
  \item \textbf{GPT-5.1 Reasoning:}
\end{itemize}

\begin{enumerate}[itemsep=-2pt, topsep=0pt, leftmargin=2.0em]
  \item \alignA{Inhibition of COX1} in renal tissues reduces synthesis of vasodilatory prostaglandins
        that normally maintain renal blood flow and glomerular filtration.
  \item \alignB{Reduced prostaglandin levels} cause renal vasoconstriction, decreased perfusion,
        and \alignC{medullary ischemia accompanied by oxidative stress}.
  \item Sustained ischemic and oxidative injury progresses to
        \alignD{proximal tubular damage and impaired renal function},
        resulting in kidney toxicity.
\end{enumerate}

\begin{itemize}[itemsep=-3pt, topsep=0pt, leftmargin=1.2em]
  \item \textbf{Predicted Toxicity:} kidney toxicity
\end{itemize}

\vspace{0.4em}
\textbf{GPT-5.1 Final Predicted Toxicity: kidney toxicity}
\vspace{0.4em}

\endgroup
\end{tcolorbox}

\captionsetup{skip=8pt, labelfont=bf}
\caption{Case study example illustrating alignment between a COX1-mediated renal toxicity AOP (Aop:177) and the GPT-5.1 reasoning path. Highlighted segments indicate correspondence between the molecular initiating event, key events, and adverse renal outcomes along the AOP cascade.}
\label{tab:aop_case_study_example2}
\end{table*}
\begin{table*}[t]
\centering
\begin{tcolorbox}[
  width=\linewidth,
  colback=gray!3,
  colframe=black,
  title=Case Study: AOP Alignment \& Reasoning Quality (o3)
]
Below is an example instance illustrating strong alignment between the \textbf{Gold AOP context (Aop:318)} and the \textbf{o3} reasoning path, demonstrating accurate reproduction of a GR-mediated hepatic steatosis pathway.

\begingroup
\setlength{\baselineskip}{0.93\baselineskip}
\vspace{0.7em}

\textbf{Sample Metadata.}
\vspace{0.4em}
\begin{itemize}[itemsep=-3pt, topsep=0pt, leftmargin=1.2em]
  \item \textbf{Index:} 54
  \item \textbf{SMILES:} \texttt{CC\#CC1(CCC2C1(CC(C3=C4CCC(=O)C=C4CCC23)C5=CC=C(C=C5)N(C)C)C)O}
  \item \textbf{AOP ID:} \texttt{Aop:318}
\end{itemize}

\vspace{0.7em}
\textbf{Gold AOP Context.}
\vspace{0.4em}
\begin{itemize}[itemsep=-3pt, topsep=0pt, leftmargin=1.2em]
  \item \textbf{AOP Title:} Glucocorticoid Receptor activation leading to hepatic steatosis
  \item \textbf{AOP:}
  \hspace*{0.8em}\textit{\alignA{Activation of Glucocorticoid Receptor}} $\rightarrow$
  \textit{\alignB{Decrease of Acyl-CoA dehydrogenases}} $\rightarrow$
  \textit{\alignC{Decreased mitochondrial fatty acid $\beta$-oxidation}} $\rightarrow$
  \textit{\alignD{Accumulation of triglyceride}} $\rightarrow$
  \textit{\alignD{Increased liver steatosis}}
\end{itemize}

\vspace{0.7em}
\textbf{o3 Reasoning Result}
\vspace{0.4em}
\begin{itemize}[itemsep=-3pt, topsep=0pt, leftmargin=1.2em]
  \item \textbf{MIE:} \alignA{GR Activation}
  \item \textbf{o3 Reasoning:}
\end{itemize}

\begin{enumerate}[itemsep=-2pt, topsep=0pt, leftmargin=2.0em]
  \item The molecule \alignA{activates the glucocorticoid receptor (GR)} in hepatocytes, altering transcription of metabolic genes.
  \item GR-driven metabolic reprogramming promotes lipid storage and functionally \alignB{reduces fatty-acid oxidation capacity},
        consistent with \alignC{decreased mitochondrial $\beta$-oxidation}.
  \item Persistent lipid accumulation  \alignD{hepatic triglyceride buildup}, progressing to \alignD{liver steatosis} and liver toxicity.
\end{enumerate}

\begin{itemize}[itemsep=-3pt, topsep=0pt, leftmargin=1.2em]
  \item \textbf{Predicted Toxicity:} liver toxicity
\end{itemize}

\vspace{0.4em}
\textbf{o3 Final Predicted Toxicity: liver toxicity}
\vspace{0.4em}

\endgroup
\end{tcolorbox}

\captionsetup{skip=8pt, labelfont=bf}
\caption{Case study example illustrating strong alignment between a GR-mediated Gold AOP (Aop:318) and the model reasoning path. Highlighted segments denote aligned molecular initiating events, key events, and adverse outcomes along the AOP cascade.}
\label{tab:aop_case_study_example3}
\end{table*}
\begin{table*}[t]
\centering
\begin{tcolorbox}[
  width=\linewidth,
  colback=gray!3,
  colframe=black,
  title=Case Study: AOP Alignment \& Reasoning Quality (Qwen3-4B-Instruct)
]
Below is an example instance illustrating transporter-mediated \textbf{kidney toxicity}, where the \textbf{Qwen3-4B-Instruct} reasoning path partially but coherently aligns with the \textbf{Gold AOP context (Aop:138)} describing OAT1 inhibition–driven renal failure.

\begingroup
\setlength{\baselineskip}{0.93\baselineskip}
\vspace{0.7em}

\textbf{Sample Metadata.}
\vspace{0.4em}
\begin{itemize}[itemsep=-3pt, topsep=0pt, leftmargin=1.2em]
  \item \textbf{Index:} 8
  \item \textbf{SMILES:} \texttt{C1=CC(=C(C=C1C2=C(C=C(C=C2)F)F)C(=O)O)O}
  \item \textbf{Model:} Qwen3-4B-Instruct
\end{itemize}

\vspace{0.7em}
\textbf{Ground Truth AOP Context:}
\vspace{0.4em}
\begin{itemize}[itemsep=-3pt, topsep=0pt, leftmargin=1.2em]
  \item \textbf{AOP ID:} \texttt{Aop:138}
  \item \textbf{AOP Title:} Organic anion transporter (OAT1) inhibition leading to renal failure and mortality
  \item \textbf{AOP:} \\
  \hspace*{0.8em}\textit{\alignA{OAT1 Inhibition}} $\rightarrow$
  \textit{\alignB{Increased blood uric acid concentration}} $\rightarrow$
  \textit{Urate deposition} $\rightarrow$
  \textit{\alignC{Renal proximal tubular necrosis}} $\rightarrow$
  \textit{Hyperkalemia} $\rightarrow$
  \textit{Renal failure / mortality}
\end{itemize}

\vspace{0.7em}
\textbf{Qwen3-4B Reasoning Result}
\vspace{0.4em}
\begin{itemize}[itemsep=-3pt, topsep=0pt, leftmargin=1.2em]
  \item \textbf{MIE:} \alignA{OAT1 Inhibition}
  \item \textbf{Qwen Reasoning:}
\end{itemize}

\begin{enumerate}[itemsep=-2pt, topsep=0pt, leftmargin=2.0em]
  \item \alignA{OAT1 inhibition} reduces the renal uptake of organic anions, including drugs and endogenous metabolites.
  \item This leads to \alignB{decreased renal clearance} and elevated plasma concentrations of potentially nephrotoxic compounds.
  \item Accumulation of these substances induces \alignC{tubular damage}, impairing nephron function and progressing to kidney injury.
\end{enumerate}

\begin{itemize}[itemsep=-3pt, topsep=0pt, leftmargin=1.2em]
  \item \textbf{Predicted Toxicity:} kidney toxicity
\end{itemize}

\vspace{0.4em}
\textbf{Qwen3-4B Final Predicted Toxicity: kidney toxicity}
\vspace{0.4em}

\endgroup
\end{tcolorbox}

\captionsetup{skip=8pt, labelfont=bf}
\caption{Case study example illustrating a transporter-mediated kidney toxicity pathway. The Qwen3-4B model reproduces the core OAT1 inhibition mechanism and downstream renal injury events defined in the Gold AOP (Aop:138), despite additional off-target reasoning blocks in other samples. Highlighted segments denote aligned molecular initiating events, key events, and adverse outcomes.}
\label{tab:aop_case_study_example4}
\end{table*}
\begin{table*}[t]
\centering
\begin{tcolorbox}[
  width=\linewidth,
  colback=gray!3,
  colframe=black,
  title=Case Study: AOP Alignment \& Reasoning Quality (ToxReason-4B-GRPO)
]

Below is an example instance illustrating ion-channel–mediated \textbf{cardiotoxicity}, where the
\textbf{ToxReason-4B-GRPO} reasoning path shows strong and coherent alignment with the
\textbf{Gold AOP context (Aop:555)} describing hERG/KCNH2 inhibition–driven electrophysiological
dysfunction leading to heart failure.

\begingroup
\setlength{\baselineskip}{0.93\baselineskip}
\vspace{0.7em}

\textbf{Sample Metadata.}
\vspace{0.4em}
\begin{itemize}[itemsep=-3pt, topsep=0pt, leftmargin=1.2em]
  \item \textbf{Index:} 28
  \item \textbf{SMILES:} \texttt{C1=CC=C2C(=C1)C=CC3=CC=CC=C3N2C(=O)N}
  \item \textbf{Model:} ToxReason-4B-GRPO
\end{itemize}

\vspace{0.7em}
\textbf{Ground Truth AOP Context:}
\vspace{0.4em}
\begin{itemize}[itemsep=-3pt, topsep=0pt, leftmargin=1.2em]
  \item \textbf{AOP ID:} \texttt{Aop:555}
  \item \textbf{AOP Title:} Inhibition, Ether-a-go-go (ERG) Voltage-Gated Potassium Channel leading to heart failure
  \item \textbf{AOP:} \\
  \hspace*{0.8em}\textit{\alignA{KCNH2 inhibition}} $\rightarrow$
  \textit{\alignB{Prolongation of action potential duration}} $\rightarrow$
  \textit{\alignC{Prolongation} \alignC{of QT interval}} $\rightarrow$
  \textit{\alignE{Increased early premature depolarizations during repolarization}} $\rightarrow$
  \textit{\alignD{Heart failure}}
\end{itemize}

\vspace{0.7em}
\textbf{ToxReason-4B-GRPO Reasoning Result}
\vspace{0.4em}
\begin{itemize}[itemsep=-3pt, topsep=0pt, leftmargin=1.2em]
  \item \textbf{MIE:} \alignA{KCNH2 inhibition}
  \item \textbf{Model Reasoning:}
\end{itemize}

\begin{enumerate}[itemsep=-2pt, topsep=0pt, leftmargin=2.0em]
  \item \alignA{KCNH2 (potassium channel hERG) inhibition} reduces potassium efflux during cardiac action potential repolarization, preventing normal membrane potential restoration.
  \item Reduced repolarization leads to \alignB{prolongation of action potential duration}.
  \item Action potential prolongation causes \alignC{QT interval prolongation}, reflecting delayed ventricular repolarization.
  \item Extended repolarization creates conditions for \alignE{early afterdepolarizations and premature} \alignE{depolarizations}, increasing arrhythmogenicity.
  \item These electrophysiological disturbances culminate in \alignD{heart failure}.
\end{enumerate}

\begin{itemize}[itemsep=-3pt, topsep=0pt, leftmargin=1.2em]
  \item \textbf{Predicted Toxicity:} cardiotoxicity
\end{itemize}

\vspace{0.4em}
\textbf{ToxReason-4B-GRPO Final Predicted Toxicity: cardiotoxicity}
\vspace{0.4em}

\endgroup
\end{tcolorbox}

\captionsetup{skip=8pt, labelfont=bf}
\caption{Case study example illustrating an ion-channel–mediated cardiotoxicity pathway.
The ToxReason-4B-GRPO model accurately reproduces the hERG/KCNH2 inhibition–driven
electrophysiological mechanism defined in the Gold AOP (Aop:555), demonstrating strong alignment
between predicted mechanistic reasoning and curated AOP knowledge. Highlighted segments denote
aligned molecular initiating events, key events, and adverse outcomes.}
\label{tab:aop_case_study_toxreason_grpo}
\end{table*}
\end{document}